\documentstyle[aps,multicol,epsf,amstex]{revtex}

\draft
\def\squote{}
\def\quote#1#2#3#4{\squote {#1,\ {\sl#2}\ {\bf#3}, #4}.\par}
\def\qquote#1#2#3#4{\squote {#1,\ {\sl#2}\ {\bf#3}, #4};}
\def\nquote#1#2#3#4{\squote {#1,\ {\sl#2}\ {\bf#3}, #4}}

\def\trans#1#2#3{[ {\sl #1} {\bf #2},\ #3 ]}
\def\prl{{\sl Phys. Rev. Lett.}\ }

\def\pr {{\sl Phys. Rev.}\ }
\def\ksi{\xi}

\begin{document}
\title{A New Spin-Orbit Induced 
Universality Class in the Quantum Hall Regime ?}
\bigskip
\author{\large Yshai Avishai and Yigal Meir}
\address{Department of Physics, Ben Gurion University, Beer Sheva 84105, ISRAEL\\and\\
The Ilse Katz Center for Meso- and Nanoscale Science and
Technology}
\maketitle
\begin{abstract}
Using heuristic arguments and numerical simulations it is argued that
the critical exponent $\nu$ describing the localization length 
divergence at the quantum Hall transition is modified in the 
presence of spin-orbit scattering with short range correlations.
The exponent is very close to $\nu=4/3$, the percolation correlation
length exponent, the prediction of a semi-classical argument. In
addition, a region of weakly localized regime, where the localization
length is exponentially large, is conjectured.
\end{abstract}
\begin{multicols}{2}
Spin-orbit scattering (SOS) is known to give rise to pronounced
effects in disordered systems. In three dimensions, the presence
of  SOS changes the universality class of the
metal-insulator transition \cite{mott}.
 More dramatically, in two dimensions, SOS of non-interacting electrons
leads to a metal-insulator transition, which does not
exist in its absence \cite{kawabata}. In the weakly localized regime,
 SOS changes localization into anti-localization \cite{hikami},
reversing the sign of the magnetoresistance,  while in the
strongly localized regime, SOS increases the localization length
(e.g. by a factor of 4 compared to its value in the absence of
SOS in quasi one dimensional systems), thus affecting the
resistance by orders of magnitude \cite{pichard}. 
The change in the universality class manifested in the level statistics
also attenuates the conductance fluctuations in the
weakly localized regime (again by a factor of 4) \cite{fluctuations}.

In spite of these remarkable effects, there have not been many
studies of the effects of SOS in the quantum Hall regime 
\cite{reviewQHE}. One
possible reason is that in the presence of a strong magnetic field,
SOS does not change the symmetry of the Hamiltonian, and thus
may not be expected to change the underlying universality class.
A counter example,  however,  exists in the strongly localized
regime, where even in the presence of a strong magnetic field,
SOS increases the localization length (again by a factor of
4 in quasi-one dimensional system, as in the absence of a field).  
Spin mixing induced by random magnetic
field was studied by Lee \cite{dkklee},
Hanna {\em et al.} \cite{hanna} (as a specific model for SOS), 
and Kagalovsky {\em et al} \cite{Victor}. The main conclusion is 
that random Zeeman term
causes splitting of the spin-degenerate quantum Hall
transition, but does not change its universality class.
The critical exponent for this kind of disorder then remains about
$2.35 \pm 0.02$, the accepted numerical value for the quantum Hall transition
\cite{huckenstein}. 
Hikami {\em et al.}\cite{zerohz} studied an electron interacting with a 
two-dimensional random magnetic field with white noise correlations, 
and demonstrated the existence of a different
universality class at $E=0$ (apparently related to non-analyticity 
of the density of states).

In this work we discuss a two-dimensional electron system in the quantum
Hall regime, subject to a random potential and random
 SOS, manifested in a spin-dependent random magnetic
field (that couples to the z-component of the electron spin), and
random spin-flip processes. The Zeeman 
splitting is assumed negligible (a specific Hamiltonian will be addressed
below). We  give heuristic arguments why one expects a change
of critical behavior of the quantum Hall transition, a change
that becomes more evident for potentials with short-range
correlations. In addition we argue that when the potential
correlation length decreases, there exists a quasi-metallic
region, where the localization length is 
exponentially large. These arguments compare very well with
numerical calculations.

The heuristic arguments are based on the semi-classical approach of 
Mil'nikov and Sokolov \cite{sokolov}, which was later applied
 to the quantum Hall effect in layered three dimensional systems\cite{meir3D}.
In the semi-classical description,
the electron follows  skipping orbit trajectories around potential
hills or valleys,  and there is a critical energy $E_c$ where the trajectory
percolates through the system \cite{classical}.
Thus, away from the critical energy
the electron is confined to a percolation cluster of typical size
$\ksi_p$, the percolation correlation length. Near threshold
$\ksi_p\sim|E_c-E|^{-\nu_p}$, where $\nu_p=4/3$ is the
two-dimensional percolation exponent. As one approaches the
transition the clusters approach each other near saddle points of
the potential energy landscape. While classically the electron
cannot move from one cluster to another,  quantum mechanically it
can tunnel through the potential barrier. If the electron energy
$E$ is close enough to the transition, the potential barrier is
close to parabolic and the tunneling probability through such a
saddle point is proportional to ${\rm exp}[-(E_c-E)]$. The number
of such saddle points through which tunneling occurs in a system
of length $L$ is typically $L/\ksi_p$. Since the transmission
coefficient is multiplicative, the conductance $\sigma$ (or the tunneling
probability) through the whole system is
\begin{equation}
\sigma \sim \left[ e^{-(E_c-E)} \right]^{L/\ksi_p} \equiv
e^{-L/\ksi} , \label{sigma2d}
\end{equation}
with $\ksi\sim (E_c-E)^{-\nu}$
 and $\nu= \nu_p + 1 = 7/3$.
The numerical estimate
$\nu=2.35\pm0.02$ \cite{huckenstein}, which is somewhat supported by
experimental data \cite{wei}, has a surprisingly excellent
agreement with the result of the above argument, especially in
view of its crudeness.

Following \cite{meir3D}, this argument can be generalized to include 
SOS.
If the spin-dependent part of the Hamiltonian is slowly
varying, one can carry out a local gauge transformation, so that
the local spin points in the direction of the local effective
random magnetic field (generated by the SOS potential)
\cite{dkklee}.  
In the adiabatic limit, where the spin-dependent potentials vary 
slowly in space, the problem separates into two independent ones
with different critical energies, split by twice the typical magnetic
field $H_{eff}$. Nonadiabaticity (short-range correlations) 
leads to mixing between these two effective
spin directions.
Consequently,  one may repeat the above argument taking into account
 the fact that   the
critical energy $E_c$ in this case is not equal to
the potential energy of the saddle-point, but is $H_{eff}$ away from
it \cite{meir3D}. Thus, the conductance $\sigma_{so}$
is
\begin{equation}
\sigma_{so} \sim \left[ e^{-H_{eff}} \right]^{L/\ksi_p} \equiv
e^{-L/\ksi_{so}} , \label{sigmaso}
\end{equation}
with $\ksi_{so}\sim (E_c-E)^{-\nu}$
 and $\nu= \nu_p  = 4/3$.

So the semi-classical argument predicts that the 
localization length  critical exponent is equal to the 
two-dimensional classical percolation exponent. 
The physical picture behind the reduction in the localization exponent
is simple: since the potential 
landscape for the opposite spin 
directions is different, then, 
due to the random effective magnetic field,
an electron approaching a saddle point may ``prefer" to flip its
spin (rather than tunnel through the saddle point), and then 
continue to propagate
semiclassically. The probability for such a Zener tunneling depends on
the potential gradients at the point, and is exponentially close to
unity  for rapidly changing potentials \cite{raikh}.
 In fact, since the tunneling
probability at the saddle point energy is
equal to $1/2$ \cite{halperin}, one
may expect that for rapidly changing potentials
there will be a region in energy where the electron will
 always ``prefer" to flip its spin as it approaches the saddle point,
and thus may cross the system classically. Quantum effects in two
dimensions will localize the electron, but the localization length
in this anomalous regime is expected to be exponentially large.

To check these predictions we use a specific, physically relevant model.
Consider the lowest two Landau levels, and denote the states in these levels
by $|n_\sigma,k>$, where $n$ is the Landau level index (0 or 1), $\sigma$
is the spin and $k$ is the momentum. 
We consider the case where, in the absence 
of disorder,  the spin-down state in the lower Landau level is degenerate
with the spin-up state in the upper one.
This may happen for electrons with a magnetic field dependent g-factor,
and seems to be relevant for composite fermions, 
 e.g. with filling factors around $\nu=3/2$ \cite{tsui},  or at filling
  factors $\nu=2/3$ and $\nu=4/5$ \cite{smet}. 
We consider only the subspace of these two degenerate Landau levels
(i.e. assume that the Landau level splitting is much larger than
the disorder potential and the SOS potential).
The disorder Hamiltonian ${\cal H}_1$ is of the form
\begin{equation}
{\cal H}_1 = V(x,y,z) + 
\alpha {\bf\sigma}\cdot{\bf \nabla}V(x,y,z)\times{\bf \Pi}, 
\label{Hso}
\end{equation} 
where $V(x,y,z)$ is the {\sl three-dimensional} 
disorder potential, $\Pi\equiv {\bf p - eA/c}$
is the electron kinetic momentum,  and $\alpha$ determines the
strength of SOS. As the momentum is constrained
to two-dimensions, taking the  limit $z\rightarrow 0$ yields, 
\begin{equation}
{\cal H}_1 = V(x,y) + 
\alpha  V_z(x,y)\left(\sigma_y\Pi_x-\sigma_x\Pi_y\right),
\label{Hso1}
\end{equation}
with $V_z\equiv\partial V/\partial z$.
Since the operator $\Pi$ operating on a state $|0_\sigma,k>$ yields
the state $|1_\sigma,k>$,  the matrix form of above Hamiltonian, 
$<n_\sigma,k|{\cal H}_1|n'_{\sigma'},k'>$ reads,
\begin{equation}
\begin{pmatrix}<0k|V|0k'> & <1k|V_{so}|1k'>\\
 <1k'|V_{so}|1k> & <1k|V|1k'>
\end{pmatrix},
\label{H}
\end{equation}
where the Landau level index now implicitly carries also the spin
quantum number,  and $V_{so}\equiv \alpha V_z$. The 
regular random potential then plays the role of  an effective random
magnetic field (which couples to the $z$-component
of the generalized spin),  while the SOS
term allows random spin-flips. In order to investigate the effects
of the latter term, the random potentials $V$ and $V_{so}$ are 
assumed independent. We fix the parameters of $V$ and
vary those of $V_{so}$, 
namely, its strength
$V_0$ (relative to the strength of the disorder potential
which defines the energy unit), and its correlation distance $\lambda$,
\begin{equation}
<V_{so}(x,y)V_{so}(x',y')> = V_0^2 f(x-x') f(y-y') ,
\end{equation}
with
$f(x) = (2\pi\lambda)^{-1/2} e^{-{{x^2}/{2\lambda}}}$.
The corresponding correlation distance for the disorder potential $V(x,y)$
was taken to be unity (all lengths are  
expressed in units of the magnetic length).

\vskip  -1.0 truecm
\begin{center}
\leavevmode \epsfxsize=3.5in
\epsfbox{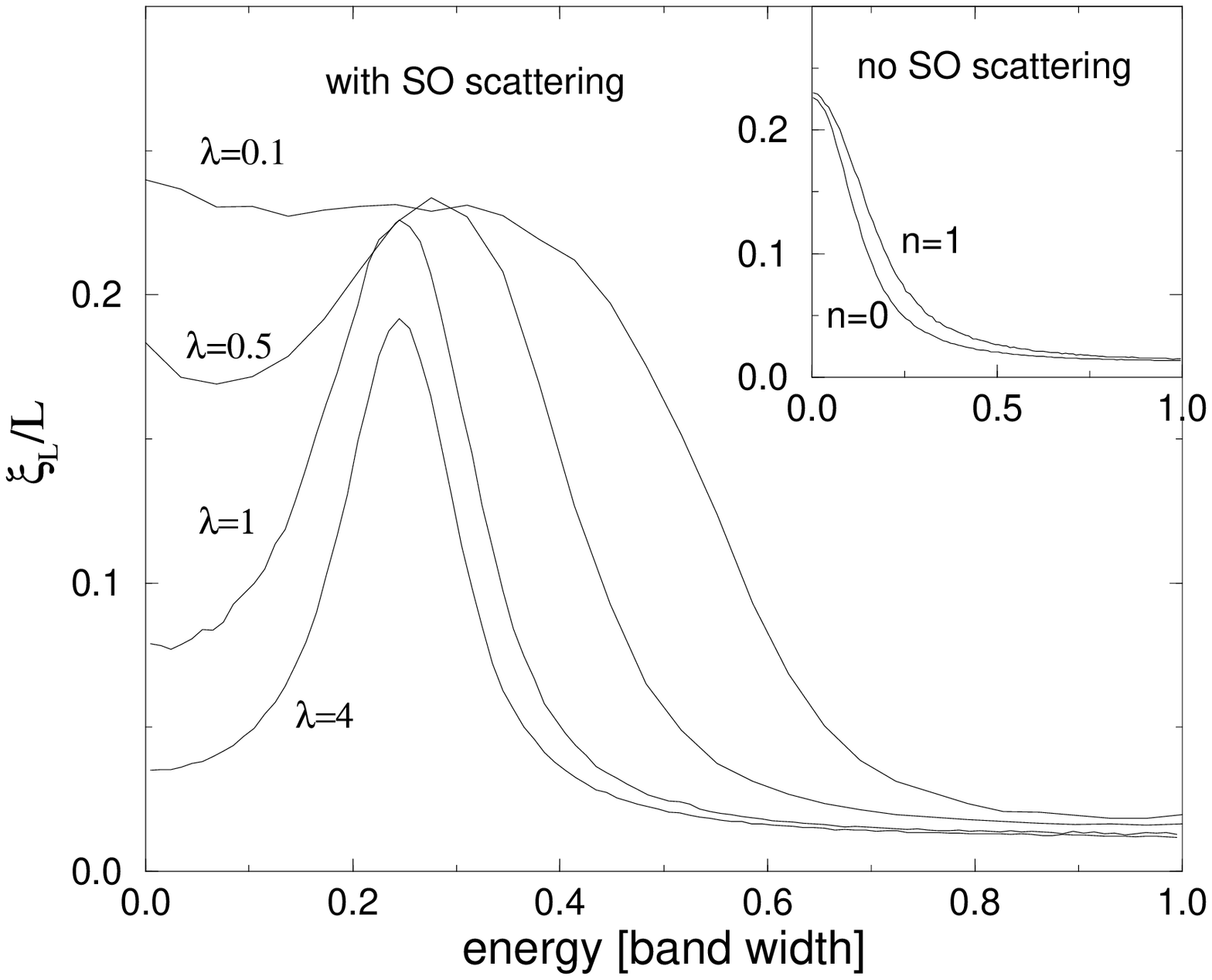}
\end{center}
\begin{small}
\vskip -0.5 truecm
Fig. 1. The finite-size localization length $\ksi_L$ scaled 
by the system size $L$, for $L=80$,
as a function of energy,
 for different values of $\lambda$, the correlation length
of the SOS potential. SOS causes 
a splitting of the quantum Hall transition, an increase in the localization
 length and, for small $\lambda$, a quasi-metallic (weakly localized)
 regime between the two
critical points (the graph is symmetric around $E=0$ and thus only positive
$E$'s are plotted). In the inset we show $\ksi_L/L$ in the absence
of SOS in the lowest two Landau levels. $\ksi_L$ is 
 maximal at $E=0$, the critical energy in this case. 
\end{small}
\vskip 0.5 truecm

Given the parameters $V_0$ and $\lambda$ we generate 
and diagonalize an ensemble of
random Hamiltonians of the form (\ref{H}) in the space of the lowest
two Landau levels, using the Landau gauge and periodic boundary 
conditions in the x-direction, for
squares of different sizes $L=$40, 60 and 80. The localization length
of a specific eigenstate $\Psi$ is determined by \cite{cardy}
\begin{equation}
\ksi_L^2\left[\Psi\right] \propto \int y^2 |\Psi(x,y)|^2 dx dy - 
\left(\int y |\Psi(x,y)|^2 dx dy\right)^2 .
\label{ksi}
\end{equation}
By dividing the energy spectrum into bins and averaging over many
disorder realizations,  we are able to obtain the energy dependence
of $\ksi_{L}(E)$.
In Fig.~1 we plot $\ksi_L(E)/L$ for several values of $\lambda$, 
for $L=80$ and $V_0=4$. 
The immediate conclusions one can draw from the figure are the
following: (1) The localization length, 
which in the absence of SOS
 was maximal at $E=0$ (see inset),  now has a maximum at
two energies, 
$E=\pm E_c$ (we show only positive energies - the graph is symmetric
around $E=0$). This leads to a splitting of the quantum Hall 
transition.
(2) The localization length increases with decreasing
$\lambda$ on both sides of the critical point,  in accordance with the
 above arguments. (3) At small $\lambda$, the localization length in the
 region between the two critical points is constant and of the order of
 the size of the system,  again in accordance with the semi-classical
 arguments.
\vskip  -1.3 truecm
\begin{center}
\leavevmode \epsfxsize=3.5in
\epsfbox{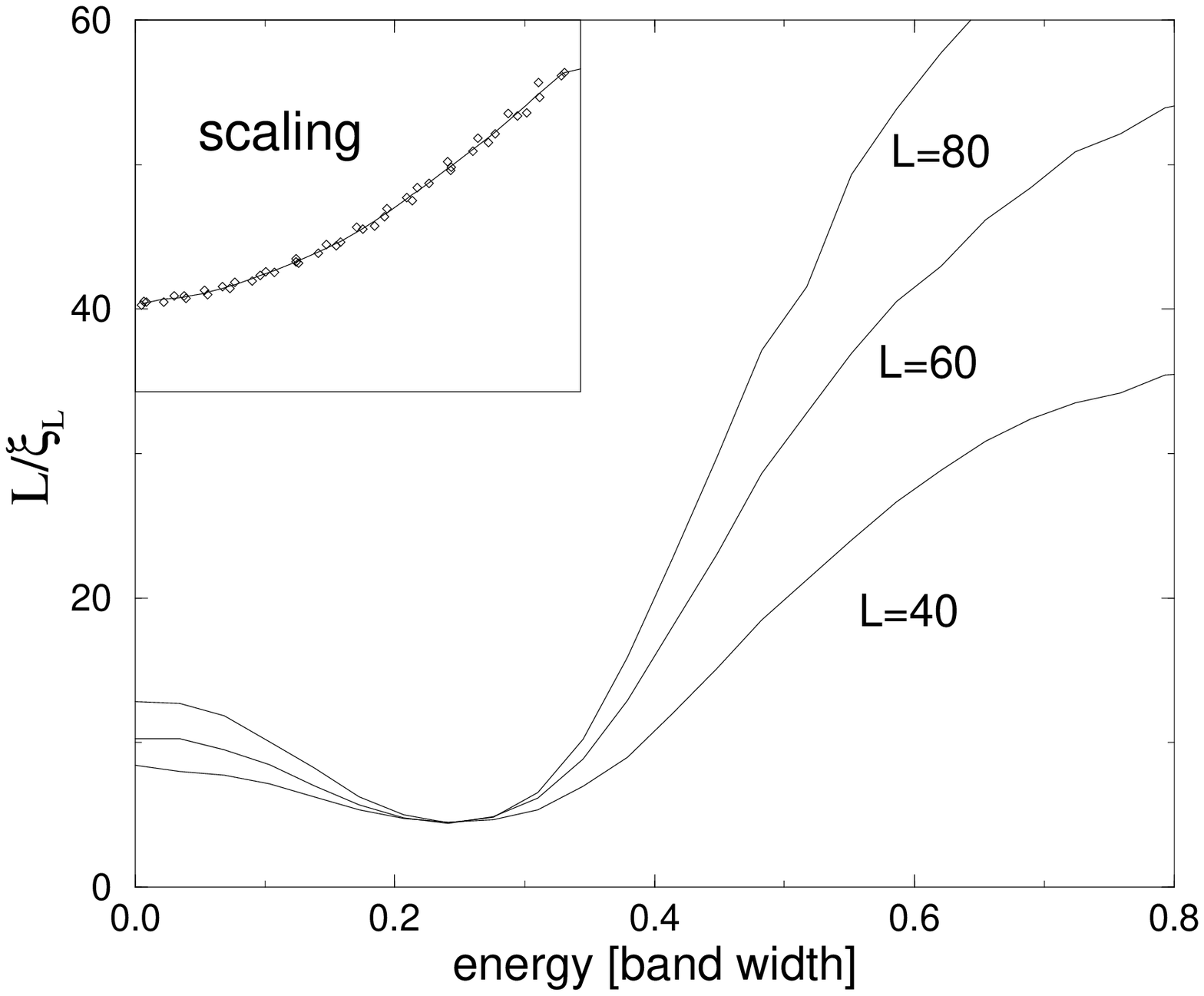}
\end{center}
\begin{small}
\vskip -1.0 truecm
Fig. 2. The inverse finite-size localization length $\ksi_L$ scaled
 by the system size $L$,
as a function of energy for different values of $L$, for $\lambda=1$.
 The value of $E$ where the 
curves meet determines the critical energy. By scaling the curves
near the critical energy (inset), one finds $\ksi(E)$, from which
the critical exponent is determined.
\end{small}
\vskip 0.5 truecm

In order to determine whether the change in the localization length is 
simply due to a numerical prefactor,  or rather,
to a different critical behavior, 
we carry out the usual scaling analysis - evaluate $\ksi_L(E)$
for different $L$'s, and collapse
all the data onto a single plot after scaling the
system size by $\ksi$,  (the $L\rightarrow\infty$ localization length), 
and set $L/\ksi_L(E) = F \left[L/\ksi(E)\right]$.
In Fig.~2 we plot $L/\ksi_L(E)$ for different $L$'s, for $\lambda=1$.
At the critical  energy $E=E_c$,  $L/\ksi_L(E)$ does not change, 
and thus $\ksi$ diverges as $E\rightarrow E_c$. The scaling of
all curves onto a single plot (Fig. 2, inset) determines
$\ksi(E)$,  and by fitting $\ksi(E)\sim (E-E_c)^{-\nu}$,  we obtain
the critical exponent $\nu$.

Fig.~3 depicts the
 derived best values of $\nu$ as a function of $\lambda$. We find that 
 for small $\lambda$ the critical exponent is indeed very close to the
 expected value from the semi-classical argument,  $\nu=\nu_p=4/3$. As
 expected,  when $\lambda$ increases $\nu$ eventually increases and 
  approaches its regular quantum Hall value. We find that for very
large $\lambda$ (not plotted) the critical exponent is 
very close to 
the regular quantum Hall exponent. Interestingly, for small $\lambda$, if
we try to scale $\ksi_L(E)$ for a large range of $E$ around $E_c$, we
find a larger critical exponent, closer to the quantum Hall one. This
may indicate that the system flows towards its low energy critical
behavior by passing close to the quantum Hall critical point. We cannot,
 however, based on our numerical procedure, determine the full phase
diagram of the quantum Hall effect in the presence of SOS.

\vskip  -0.5 truecm
\begin{center}
\leavevmode \epsfxsize=3.5in
\epsfbox{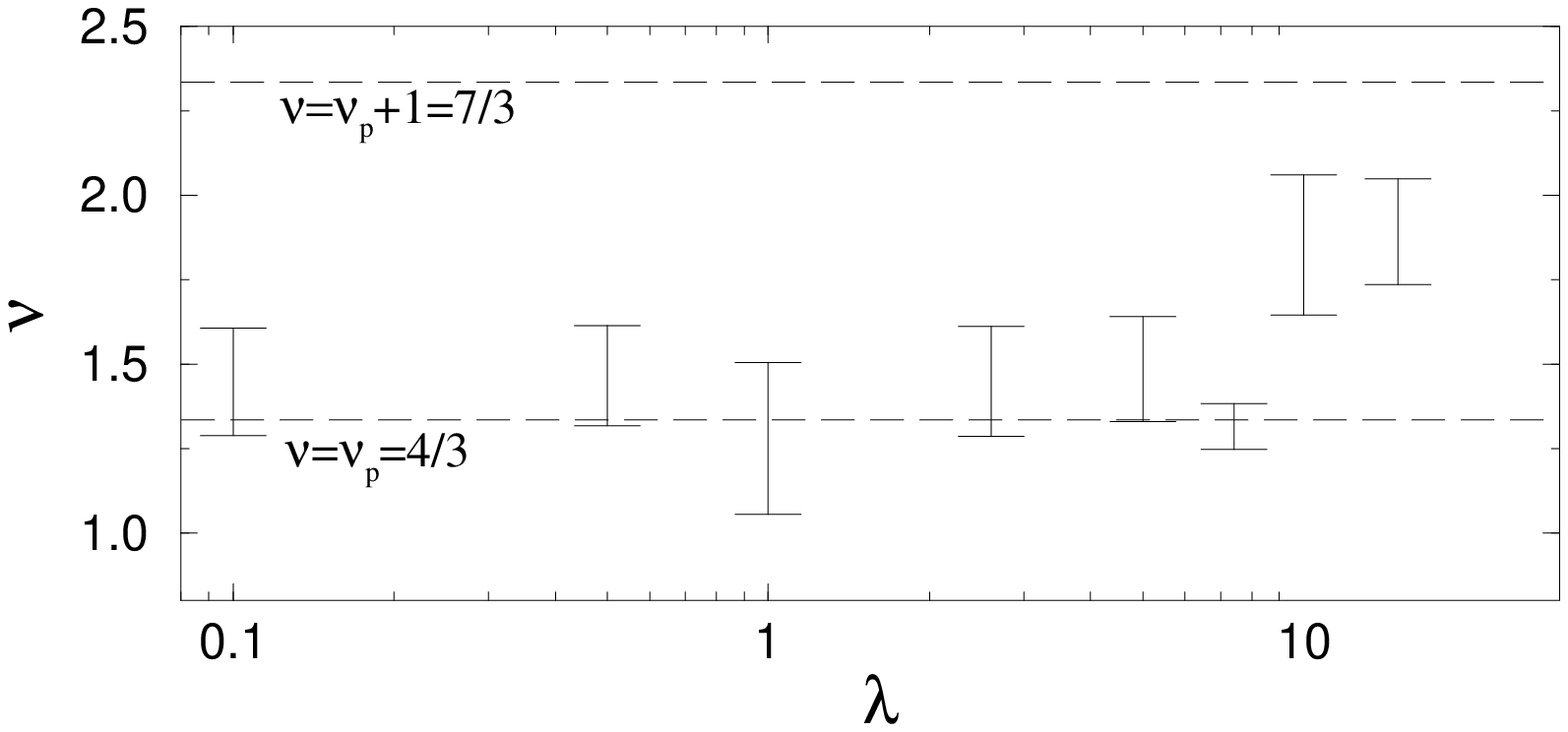}
\end{center}
\begin{small}
\vskip -0.2 truecm
Fig. 3. The derived values of $\nu$, the localization length critical
exponent, as a function of $\lambda$, the correlation length of the 
SOS potential. For small $\lambda$ the exponent is
close to $\nu_p=4/3$, the percolation correlation length exponent, in
 agreement with the prediction of the semi-classical theory. This theory
also predicts that the critical behavior will crossover to the regular
quantum Hall transition ($\nu\simeq 7/3$) for large $\lambda$, in agreement
with the numerical results.
\end{small}
\vskip 0.5 truecm

It is interesting to compare our results with previous works. 
Lee \cite{dkklee} and Hanna {\em et al.}\cite{hanna} 
 studied an Hamiltonian with a 
spin-dependent term ${\bf H}({\bf r})\cdot{\bf S}$, 
in which ${\bf H}({\bf r})$ is a random
field that couples to the electron spin ${\bf S}$. Their conclusion is
that, at least for random field which varies smoothly in space,
the quantum Hall transition splits, but the critical behavior remains
unchanged. Hanna {\em et al.} also noted that
 for random field with
short-range (white noise) correlations, the conductance is peaked at $E=0$, 
similarly to our curve for $\ksi_L/L$ for $\lambda=0.5$ in Fig.~1.
Their interpretation is that the critical energy may have shifted 
close to $E=0$, but the critical behavior remains that of
the regular quantum Hall effect. If one assumes, however, a single
critical point at $E=0$, one finds a critical exponent {\sl larger}
than the quantum Hall one (which may explain the experimental
observations for spin-degenerate  Landau levels \cite{wei}).
Indeed we checked that within the  
${\bf H}({\bf r})\cdot{\bf S}$  model, even if the
correlation length of the random magnetic field decreases, the
critical behavior remains quantum-Hall-like 
($\nu\simeq 2.35$).
The difference between that model and ours  might seem
surprising,
since the model we use (\ref{H}) looks very much like a random field.
However, the distinction becomes apparent in Fig.~4,
where we plot the density of states and $\ksi_L(E)/L$ for the two models.
While for the ${\bf H}({\bf r})\cdot{\bf S}$ model 
the density of states splits into two
peaks, indicating a splitting of the spin-degenerate Landau level into 
two independent Landau levels, in our model the density of states remains
peaked at $E=0$ (even though the critical points move away from it),
and thus the two effective spin-directions are still strongly mixed.
The motivation to study the ${\bf H}({\bf r})\cdot{\bf S}$ 
model stems from the fact
that the magnetic field breaks the simplectic symmetry of the spin-orbit
Hamiltonian, and thus it was assumed enough to study an Hamiltonian with
a unitary symmetry. Our model, however, still obeys the full symmetry
one expects for SOS in the presence of a magnetic field --
time-reversal symmetry followed by reversing the (uniform) 
magnetic field. Thus,
it is expected that the two models belong to different 
universality classes. A similar argument has been successfully applied
to the change in localization length in the strongly localized regime
\cite{pichard} where it was shown that an application of SOS
 changes the scaling of the localization length, even
in the presence of a strong magnetic field, though the universality
class might have been expected to remain the same. 
\vskip  -1.8 truecm
\begin{center}
\leavevmode \epsfxsize=3.5in
\epsfbox{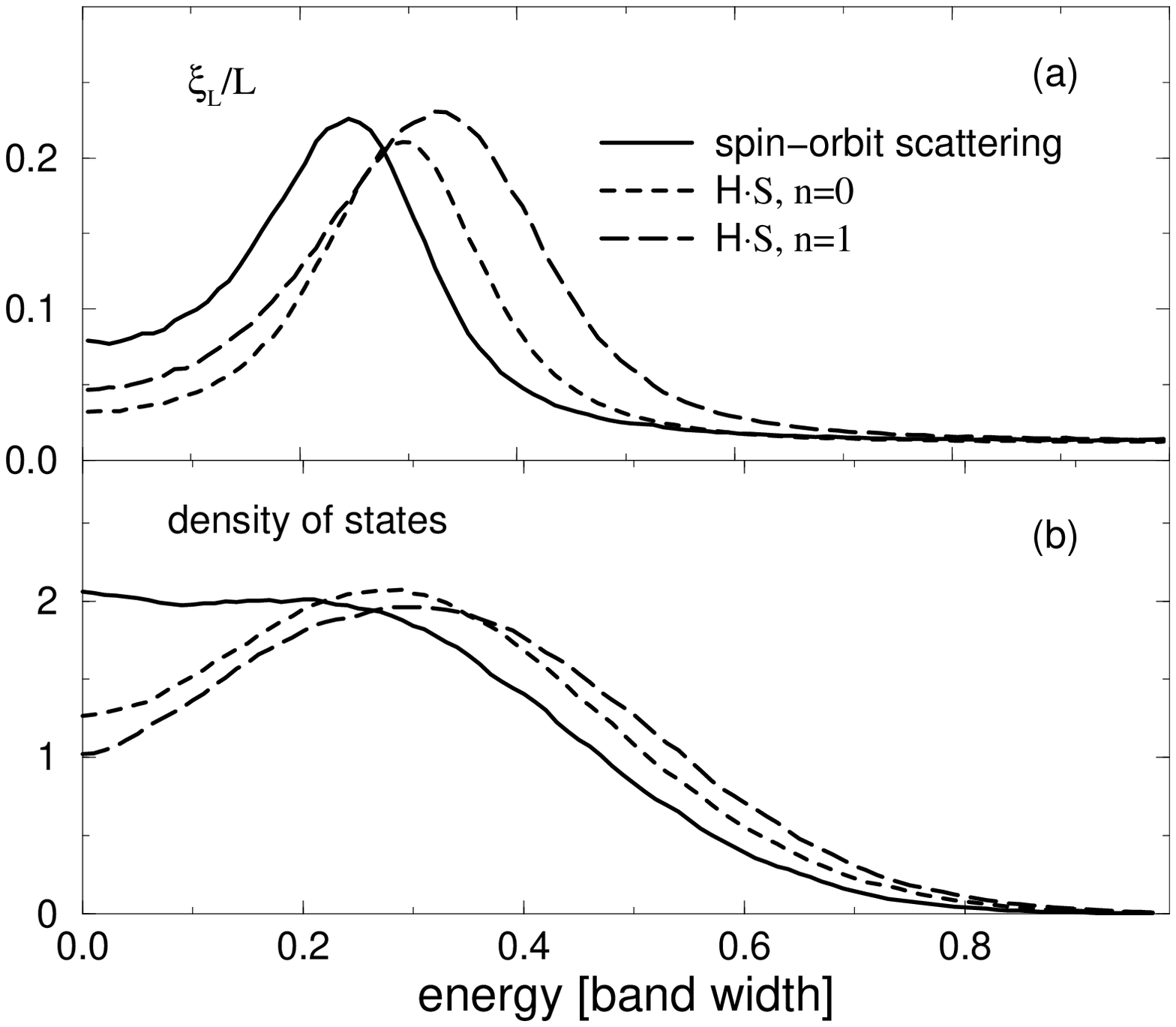}
\end{center}
\begin{small}
\vskip -0.5 truecm
Fig. 4. Comparison of our model (\ref{H}) to the ${\bf H}({\bf r})
\cdot{\bf S}$ model,
(a) $\ksi_L(E)/L$, and (b) density of states, 
for $L=80$ and $\lambda=1$. While the
critical energies in both models split from $E=0$, the density of
 states in our model remains peaked at $E=0$, in contrast with the
split density of states for the ${\bf H}({\bf r})\cdot{\bf S}$ 
model, allowing for
stronger mixing of the spins and a possible change in the critical behavior.
\end{small}
\vskip 0.5 truecm

To conclude, we have presented arguments and
demonstrated numerically that spin-orbit scattering
in the quantum Hall regime may alter the critical 
behavior for potentials with short range correlations. 
The calculated critical exponent agrees very well with
the percolation correlation length exponent, a value predicted by
semi-classical arguments.

This work is supported
 by DIP, BSF and ISF funds.

\end{multicols}

\begin{references}
\bibitem{mott}
For a review, see,  e.g.,  {\sl Metal-Insulator Transitions}, 
N. F. Mott (Taylor \& Francis,  London,  1990).

\bibitem{kawabata}
\qquote{S. N. Evangelou and T. Ziman}{J. Phys.}{C 20}{L235 (1987)}
\quote{A. Kawabata}{J. Phys. Soc. Japan}{57}{1717 (1988)}

\bibitem{hikami}
\qquote{S. Hikami, A. I. Larkin and Y. Nagaoka}{Prog. Theor. Phys.}
{63} {707 (1980)}
\qquote{G. Bergmann}{\prl}{48}{1046 (1982)}
\quote{See also Y. Meir,  Y. Gefen and O. Entin-Wohlman}{\prl}{63}{798 (1989)}

\bibitem{pichard}
\qquote{J. L. Pichard et al.}{\prl}{65}{1812 (1990)}
\qquote{Y. Meir et al.}{\prl}{66}{1517 (1991)}
\qquote{E. Medina and M. Kardar}{\prl}{66}{3187 (1991)}
\quote{Y. Meir and O. Entin-Wohlman}{\prl}{70}{1988 (1993)}

\bibitem{fluctuations}
\nquote{B.~L. Altshuler and B.~I. Shklovskii}{Zh. Eksp. Teor. Fiz.}{91}
{220 (1986)} \trans{Sov. Phys. JETP} {64}{127 (1986)};
\quote{N. Zanon and J.-L. Pichard}{J. Phys. (Paris)}{49}{907 (1988)}

\bibitem{reviewQHE}
For a review see, e.g., M. Stone, {\sl The Quantum Hall Effect} (World
Scientific,  Singapore,  1992).

\bibitem{dkklee}
\quote{D. K. K. Lee}{\pr}{B 50}{7743 (1994)}

\bibitem{hanna}
\quote{C. B. Hanna et al.}{\pr}{B 52}{5221 (1995)}

\bibitem{Victor} 
V. Kagalovsky {\em et al}, 
Phys. Rev. B55, 7761 (1997).

\bibitem{huckenstein}
For a review of the numerical work,  see \quote{B. Huckestein}{Rev. Mod. Phys.}
{67}{357 (1995)}

\bibitem{zerohz}
\qquote{S. Hikami, M. Shirai and F. Wegner}{Nucl. Phys.}{B408} 
{415 (1993)}
\quote{K. Minakuchi and S. Hikami} {Phys. Rev.} {B53} {10898 (1996)}

\bibitem{sokolov}
\quote{G. V. Mil'nikov and I. M. Sokolov}{JETP Lett.}{48}{536 (1988)}

\bibitem{meir3D}
\quote{Y. Meir}{Phys. Rev.}{B 58}{R1762 (1998)}

\bibitem{classical}
\qquote{S. Luryi and R. F. Kazarinov}{\pr}{B 27}{1386 (1983)}
\qquote{S. A. Trugman}{\pr}{B 27}{7539 (1983)}
 \quote{R. Mehr and A. Aharony}{\pr}{B 37}{6349 (1988)}
 
\bibitem{wei}
\qquote{H. P. Wei et al.}{\prl}{61}{1294 (1988)}
\qquote{H. P. Wei et al.}{Surf. Sci.}{229}{34 (1990)}
\qquote{S. Koch et al.}{\pr}{B 43}{6828 (1991)}
F. Hohls, U. Zeitler and R.~J. Haug, cond-mat/0107412;
See, however, \nquote{V.~T. Dolgopolov, G.~V. Kravchenko and A.~A. Shashkin}
{\pr}{B 46}{13303 (1992)} and  \nquote{A.~A. Shashkin,  V.~T. Dolgopolov
and G.~V. Kravchenko}{\pr}{B 49}{14486 (1994)},  who report $\nu\simeq1$, 
and \nquote{D. Shahar et al.}{Solid Stat. Comm.}{107}{479 (1998)},  who
see no critical behavior.

\bibitem{raikh}
\quote{A. Gramada and M. E. Raikh}{\pr}{B 56}{3965 (1997)}

\bibitem{halperin}
\quote{H. A. Fertig and B. I. Halperin}{\pr}{B 36}{7969 (1987)}

\bibitem{tsui}
\quote{R. R. Du et al.}{\prl}{75}{3926 (1995)}

\bibitem{smet}
\qquote{S. Kronm\"uller et al.}{\prl}{81}{2526 (1998)}{\bf 82}, 4070 (1999);
\quote{J. H. Smet et al.}{\prl}{86}{2412 (2000)}

\bibitem{cardy}J. L.~Cardy (ed.), {\sl Finite Size Scaling} (North Holland,
Amsterdam, 1988); V.~Privman (ed.),  {\sl Finite Size Scaling and Numerical
Simulations of Statistical Systems} (World Scientific,  Singapore, 1990).

\end{references}
\end{document}